\documentclass[twocolumn,prl,superscriptaddress,showpacs]{revtex4}
\usepackage{graphicx}
\usepackage{times}
\usepackage{amsmath}

\newcommand{\ph}{\varphi}
\newcommand{\eps}{\varepsilon}

\begin{document}

\title{Interacting anyons in topological quantum liquids: The golden chain}

\author{Adrian Feiguin}
\affiliation{Microsoft Research, Station Q, University of California, Santa Barbara, CA 93106}
\author{Simon Trebst}
\affiliation{Microsoft Research, Station Q, University of California, Santa Barbara, CA 93106}
\author{Andreas W.~W.~Ludwig}
\affiliation{Physics Department, University of California, Santa Barbara, CA 93106}
\affiliation{Kavli Institute for Theoretical Physics, University of California, Santa Barbara, CA 93106}
\author{Matthias~Troyer}
\affiliation{Theoretische Physik, Eidgen\"ossische Technische Hochschule Z\"urich, 8093 Z\"urich, Switzerland}
\author{Alexei Kitaev}
\affiliation{Microsoft Research, Station Q, University of California, Santa Barbara, CA 93106}
\affiliation{Institute for Quantum Information, California Institute of Technology, Pasadena, CA 91125}
\author{Zhenghan Wang}
\affiliation{Microsoft Research, Station Q, University of California, Santa Barbara, CA 93106}
\author{Michael H.~Freedman}
\affiliation{Microsoft Research, Station Q, University of California, Santa Barbara, CA 93106}

\date{\today}

\begin{abstract}
We discuss generalizations of quantum spin Hamiltonians using anyonic degrees 
of freedom. The simplest model for interacting anyons energetically favors 
neighboring anyons to fuse into the trivial (`identity')  channel, 
similar to the quantum Heisenberg model favoring neighboring spins to form spin singlets.
Numerical simulations of a chain of Fibonacci anyons show that the model is critical 
with a dynamical critical exponent $z=1$, and described by a
two-dimensional (2D) conformal field theory with central charge $c=7/10$. 
An exact mapping of the anyonic chain onto the 2D tricritical Ising model is given
using the restricted-solid-on-solid (RSOS) representation of the Temperley-Lieb algebra.
The gaplessness of the chain is shown to have topological origin.
\end{abstract}



\maketitle


\paragraph{Introduction }
Non-Abelian anyons are exotic particles expected to exist in certain
fractional quantum Hall (FQH) states \cite{MooreRead,ReadRezayi}. A set of
several anyons supports very robust collective states that are degenerate to
exponential precision; such states can potentially be used as quantum memory
and for quantum computation~\cite{ToricCode}. However, this degeneracy can be
lifted by a short-range interaction if the anyons are very close to each
other.  As a first step towards understanding interacting anyons, we
describe a simple, exactly solvable model that is an anyonic analogue of the
quantum Heisenberg chain.
 
We start by considering the well-known Moore-Read state \cite{MooreRead},
a candidate state, exhibiting non-Abelian statistics,
for the topological nature of FQH liquids at filling fraction $\nu=5/2$.
It has two important types of excitations: quasiholes with electric charge
$e/4$ and neutral fermions. Quasiholes may be trapped by an impurity potential
while the fermions can still tunnel between them \cite{ReadLudwig}. 
For a one-dimensional (1D) array of trapped quasiholes, the Hamiltonian can be 
described in terms of free Majorana fermions on a lattice, which is in turn equivalent
to the 1D transverse field Ising model at the quantum phase transition point. The
more interesting model discussed here is based on so-called `Fibonacci
anyons', which represent the non-Abelian part of the quasiparticle statistics
in the $k=3$, $Z_k$-parafermion
state \cite{ReadRezayi}, an effective theory for
FQH liquids at filling fraction $\nu=12/5$ \cite{TwelveFifth}. 
Even without  parameter fine-tuning,  these 1D anyonic arrays
will be shown to exhibit gapless excitations due to topological symmetry.

\begin{figure}[b]
  \includegraphics[width=68mm]{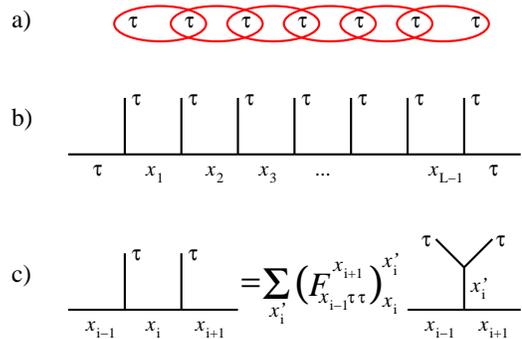}
  \caption{
    a)~Illustration of the Fibonacci chain with $L$ $\tau$-anyons. 
    b)~The fusion path. 
     c)~Definition of the $F$-matrix.
  }
  \label{Fig:FibonacciChain}
\end{figure}

\paragraph{Model }
Our model describes pairwise interactions within an array of $L$ anyons, for instance 
along a chain as shown in Fig.~\ref{Fig:FibonacciChain}a). In the Fibonacci theory 
there are only two types of particles: the Fibonacci anyon, denoted by $\tau$, and the
trivial particle denoted by 1 with a fusion rule $\tau \times \tau =
1+\tau$. We refer to the label $1$ or $\tau$ as the topological charge.  When
two neighboring anyons interact, indicated in the figure by the ellipses, they
can either fuse in the trivial channel, annihilating each other, or in the
nontrivial one, becoming a single $\tau$-anyon \cite{Preskill}.  We
define our model by assigning an energy gain if they fuse
along the trivial channel. This is an anyonic analogue
of the spin-1/2 quantum Heisenberg antiferromagnet, which 
assigns an energy gain to two neighboring spin-1/2 fusing into a spin-0
singlet as compared to a spin-1 triplet.
 
To define the Hilbert space of $\tau$-anyons we consider the
tree-like fusion diagram in Fig. \ref{Fig:FibonacciChain}b).
The basis corresponds to all admissible labelings $|x_1,x_2,\ldots\rangle$ of
the links, with $x_i=1$ or $\tau$. Each label represents the combined
topological charge of the particles left to a given point.  Not all
possible values $(x_1,x_2,\ldots)$ represent allowed basis states due to the
fusion rules: a $1$ must always be preceded and followed by a $\tau$, since
the fusion of a $1$ and a $\tau$ always gives a $\tau$. This reduces the
dimension of the Hilbert space of the open chain (with $\tau$-labels at the 
boundary) 
to the Fibonnacci sequence
dim$_L = F_{L+1}$, and for the periodic chain dim$_L = F_{L-1}+F_{L+1}$. For
large $L$ it is well-known that these numbers grow at a rate $\dim_L \propto
\ph ^L$, where $\ph = (1+\sqrt{5})/2$ is the golden ratio.  This Hilbert space
has no natural decomposition in the form of a tensor product of single-site
states, in contrast to $SU(2)$ quantum spin chains. 

In order to generate a local Hamiltonian ${\bf H}=\sum_i {\bf H}_i$
assigning an energy to the fusion of two neighboring 
$\tau$-anyons we use the so-called $F$-matrix to transform the local basis as shown in 
Fig. \ref{Fig:FibonacciChain}c). In the transformed basis the state $x_i'$ 
corresponds to the fusion of the two anyons. The Hamiltonian is then 
defined by assigning an energy $E_{\tau}=0$ for $x_i'=\tau$, and $E_1=-1$ 
for $x_i'=1$. 
The resulting  local terms ${\bf H}_i$ contain three-body interactions 
in the link basis,
\begin{eqnarray}
&& {\bf H}_i | x_{i-1} x_i x_{i+1} \rangle = 
\sum_{{x'}_i=1, \tau} 
({\bf H}_i)_{x_i}^{{x'}_i} \ \ |x_{i-1} {x'}_i x_{i+1} \rangle
\label{Eq:Hamiltonian}
\\
&& {\rm with} \quad
{(\bf H}_i)_{x_i}^{{x'}_i} := -{ ( F^{x_{i+1}}_{x_{i-1} \tau \tau} )}^{1}_{x_i} 
{ ( F^{x_{i+1}}_{x_{i-1} \tau \tau})}^{1}_{{x'}_i} \,.
\nonumber
\end{eqnarray}
It is diagonal in the subspace 
$\{| x_{i-1} x_i x_{i+1} \rangle \}=$ 
$\{|1 \tau 1\rangle,|1\tau \tau \rangle,|\tau \tau 1 \rangle \}$, 
${\bf H}_i ={\rm diag}\{-1,0,0\}$, where the $F$-matrix is
a number due to the constraints arising from the fusion rules.
For the case $x_{i-1}=x_{i+1}=\tau$, the $F$-matrix and the corresponding Hamiltonian are
the following $2\times 2$-matrices
($x_i, x_i' \in $ $\{1, \tau \}$)
\begin{equation}
 {\bf  F}_{\tau\tau\tau}^{\tau} = \begin{pmatrix} \ph^{-1} & {\ph}^{-1/2} \\
                                     {\ph}^{-1/2} & -\ph^{-1}
          \end{pmatrix} 
  \,,\quad
  {\bf H}_i = - \begin{pmatrix} \ph^{-2} & {\ph}^{-3/2} \\
                                     {\ph}^{-3/2} & \ph^{-1}
          \end{pmatrix} 
\,.
  \label{Eq:FMatrix}
\end{equation}
Looking at the matrix form of the Hamiltonian, it can be written in terms of
standard Pauli 
matrices:
\begin{eqnarray}
\nonumber
{\bf H}_i & = &
\left(n_{i-1}+n_{i+1} -1\right)
\\ \nonumber
&& - n_{i-1}n_{i+1} \left(  \ph^{-3/2}\sigma^x_i +\ph^{-3}n_i 
+1+\ph^{-2} \right) \,,
\end{eqnarray}
where
 the sum runs over the links of the chain. 
In this expression, the operators $n_i$ count the $\tau$-particle occupation on link $i$,
$n_i=\frac{1}{2}(1-\sigma^z_i) = 0,1$,
and the Hamiltonian ${\bf H}$
acts on the constrained Hilbert space defined above.

\begin{figure}[b]
  \includegraphics[width=86mm]{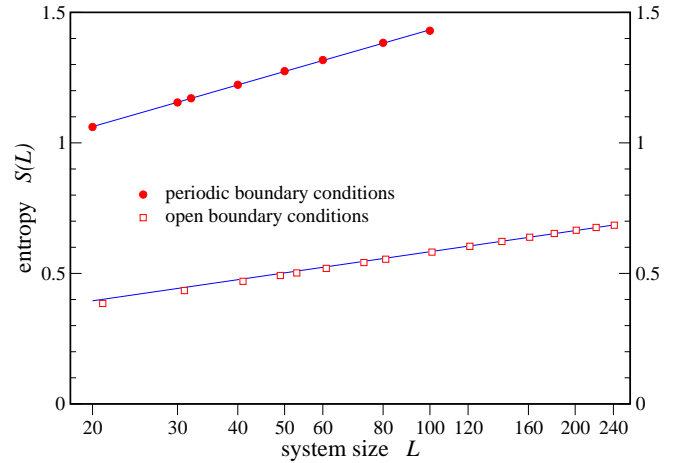}
  \caption{
   (Color online)
   Entropy scaling for interacting Fibonacci anyons arranged along an open 
    (open squares) or periodic chain (closed circles) versus the system size $L$.
    Logarithmic fits (solid lines) give central charge estimates of $c_{\rm PBC}=0.701 \pm 0.001$
    and $c_{\rm OBC}=0.70 \pm 0.01$ respectively,
    where for the open boundary conditions only the values for the 5 largest systems 
    have been taken into account due to large finite-size effects.
  } 
  \label{Fig:EntropyScaling}
\end{figure}

\paragraph{Entropy scaling and central charge  }

We simulated this model numerically, and calculated the
finite-size
 excitation gap by exact diagonalization 
for chains of up to $L=32$ sites. A finite-size analysis shows that the gap
vanishes linearly in $1/L$, indicative of a
critical
 model with dynamical critical exponent $z=1$ 
described by a conformal field theory. 
The central charge $c$ of a CFT can be calculated from the finite-size scaling of the entanglement entropy. For two subsystems with equal size $L/2$ on systems with periodic (PBC) and open boundary conditions (OBC) the entanglement entropy scales as \cite{entanglement}
\begin{equation}
 S_{\rm PBC} (L) \propto \frac{c}{3} \log{L} \quad {\rm and} \quad
 S_{\rm OBC} (L) \propto  \frac{c}{6} \log{L}\,.
  \label{Eq:EntropyScaling}
\end{equation}
The density matrix renormalization group method (DMRG) \cite{DMRG,DMRGdetails} 
provides a natural framework for calculating these quantities. Fits of our numerical results 
according to Eqs.~(\ref{Eq:EntropyScaling}) 
shown in Fig. \ref{Fig:EntropyScaling}, 
give central charge estimates of $c_{\rm PBC}=0.701 \pm 0.001$ and 
$c_{\rm OBC}=0.70 \pm 0.01$ respectively.
Since possible (unitary) CFTs in the vicinity of these 
estimates 
have central
 charges\cite{FriedanQuiShenker}
$1/2$, $7/10$ or $4/5$ we can unambiguously conclude that our results are consistent 
only with central charge $c=7/10$.


\paragraph{Mapping and exact solution }
We now proceed to derive these results exactly.
By construction the local contribution to the Hamiltonian ${1\over \varphi} {\bf X}_i = -{\bf H}_i$
is a projector
onto the trivial particle.
One can then verify that the operators 
${\bf X}_i$
form a representation of the
Temperley-Lieb algebra \cite{TemperleyLieb}
\begin{eqnarray}
\nonumber
({\bf X}_i)^2  &=& d \ {\bf X}_i,\\
\nonumber
 {\bf X}_i {\bf X}_{i\pm 1} {\bf X}_i &=& {\bf X}_i,\\
\label{TemperleyLieb}
 [{\bf X}_i, {\bf X}_j ] &=& 0  \ \ {\rm for} \ \ |i-j| \geq 2 \,,
\end{eqnarray}
where the `d-isotopy'  parameter 
equals the golden ratio, $d=\varphi$.
This representation can be seen to be identical to
the standard Temperley-Lieb algebra representation
associated with $SU(2)_k$ at level $k=3$.
For arbitrary  $k$,
the latter contains $k+1$ anyon species
labelled by $j=0, 1/2, 1, ..., k/2$,
satisfying the fusion rules of $SU(2)_k$ \cite{FusionRulesofSuTwoLevelK}.
The operators ${\bf e}_i$ defined by
\begin{eqnarray}
\nonumber
{\bf e}[i] | j_{i-1} j_i j_{i+1} \rangle 
&=& 
\sum_{{j'}_i}
\left({\bf e}[i]^{j_{i+1}}_{j_{i-1}}\right)_{j_i}^{{j'}_i}
  |j_{i-1} {j'}_i j_{i+1} \rangle
\\
{\rm and} \quad
\left({\bf e}[i]^{j_{i+1}}_{j_{i-1}}\right)_{j_i}^{{j'}_i}
&=&
\delta_{j_{i-1},j_{i+1}}
\ \ 
\sqrt{
S^0_{j_i} S^0_{{j'}_i}
\over
S^0_{j_{i-1}} S^0_{{j}_{i+1}}
}
\label{Defei}
\end{eqnarray}
are known \cite{RefTemperleyLiebForSuTwoLevelK}
to form a representation of the Temperley-Lieb algebra
(\ref{TemperleyLieb}) for any value of $k$,
where $|j_i-j_{i+1}|=1/2$ and
$S^{j'}_j := \sqrt{2 \over (k+2)} \ 
\sin [ \pi {(2j+1) (2 {j'}+1)\over k+2} ]$\cite{FootnoteModularSmatrix}.

\begin{figure}[b]
  \includegraphics[width=86mm]{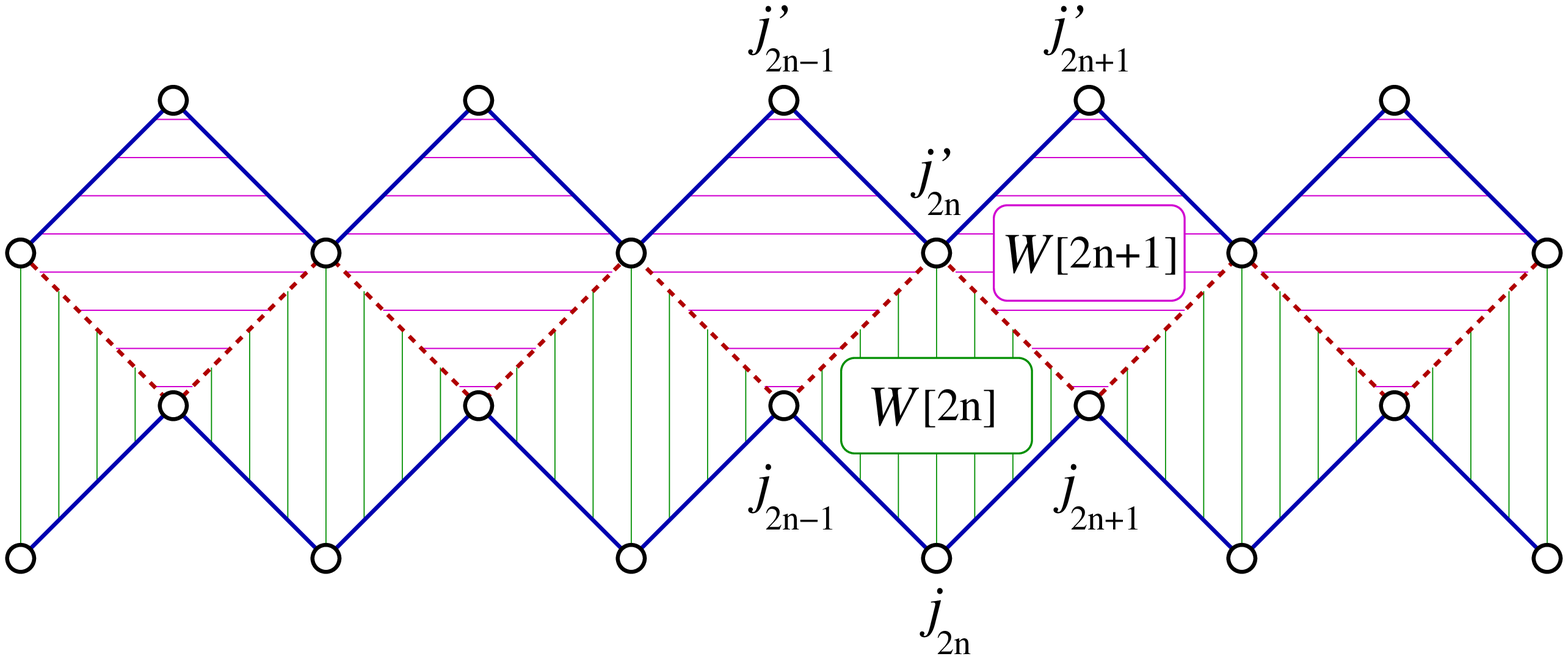}
  \caption{
  Transfer matrix of the RSOS model.
   }
  \label{Fig:TransferMatrix}
\end{figure}

Our model of interacting Fibonacci anyons can be cast into
this form at $k=3$  by first mapping
$x_i =1$ $ \to j_i=0$, and $x_i =\tau $ $ \to j_i=1$,
and then
applying
the $SU(2)_3$ fusion rule
$3/2 \times j = 3/2-j$
to the even-numbered sites.
This maps any admissible labeling 
$|{\vec x}\rangle:=$
$|x_1,x_2,\ldots\rangle$ 
uniquely into
$|{\vec j}\rangle:=$
$|j_1,j_2,\ldots\rangle$ 
where for odd-numbered sites $j_{2i+1} \in \{0, 1\}$,
and for even numbered-sites $j_{2i} \in \{1/2, 3/2 \}$.
This re-labeling maps the matrix elements of ${\bf X}_i$ into those of
${\bf e}_i$ from Eq.~(\ref{Defei}).

We can now see that the Hamiltonian in Eq.~(\ref{Eq:Hamiltonian})
is that corresponding to a standard (integrable) lattice model
description of the classical 2D tricritical Ising model, 
known as the RSOS model  \cite{RSOSmodel}.  
Specifically, the two-row transfer matrix ${\bf T}:={\bf T}_2 {\bf T}_1$ 
of this lattice model, shown in Fig.~\ref{Fig:TransferMatrix}, is written in terms 
of Boltzmann weights ${\bf W}[i]$ assigned to a plaquette $i$ of the square lattice
\begin{equation}
\nonumber
{\bf T}_1 := \prod_{n} {\bf W}[2n]\,, \quad {\rm and} \quad
\ \ \ 
{\bf T}_2 := \prod_{n} {\bf W}[2n+1]
\end{equation}
with
\begin{equation}
\label{DefWi}
{\bf W}[i]^{\vec {j'}}_{\vec j}
=
\frac{\sin[\frac{\pi}{k+2}-u]}{\sin\frac{\pi}{k+2}}\,
{\bf 1}^{\vec {j'}}_{\vec j}
+ 
\frac{\sin{u}}{\sin\frac{\pi}{k+2}}\,
{\bf e}[i]^{\vec {j'}}_{\vec j} \,.
\end{equation}
The parameter $u>0$ is a measure of the lattice anisotropy,
${\bf 1}$~is the identity operator, and
\begin{equation}
{\bf e}[i]^{\vec {j'}}_{\vec j}
:= \ \ 
\label{ProjectorsRow}
 \left[\prod_{m\not = i} \delta_{j'_m,j_m}\right]
\ \ \left({\bf e}[i]^{j_{i+1}}_{j_{i-1}}\right)_{j_i}^{j'_i} \,.
\end{equation}
The Hamiltonian of the so-defined lattice model
is obtained from its transfer matrix by taking, 
as usual \cite{BaxterBook},
the extremely anisotropic limit, $u \ll 1$,
\begin{equation}
\nonumber
{\bf T} =
\exp\{ - a ({\bf H}+c_1) + O(a^2) \},
\ \ \ a = {u \varphi \over \sin[\pi/(k+2)]} \ll 1
\end{equation}
yielding $ {\bf H} =  - \sum_i \ {1\over \varphi} {\bf e}_i$ ($c_1$ is an unimportant constant).
Since the operators ${\bf X}_i$ can be identified with ${\bf e}_i$, this demonstrates 
that the Hamiltonian of the Fibonacci chain is exactly that of the corresponding 
$k=3$ RSOS  model which is a lattice description of the tricritical Ising model at its 
critical point.
The latter is a well-known (supersymmetric) CFT with central charge $c=7/10$ 
\cite{BPZ,TricriticalIsingModel}.
Analogously one obtains \cite{Huse1984} for general $k$
the $(k-1)^{\rm st}$ unitary minimal
CFT \cite{FriedanQuiShenker} 
of central charge $c=1-6/(k+1)(k+2)$.
A {\em ferromagnetically} coupled
Fibonacci chain (energetically favoring the fusion along the $\tau$-channel)
is described by the critical 3-state Potts model with $c=4/5$ 
and, for general $k$,
by the critical $Z_k$-parafermion  
CFT \cite{uReversedSign,RSOSmodel,Huse1984} with 
central charge $c=2(k-1)/(k+2)$.
 

\begin{figure}[t]
  \includegraphics[width=86mm]{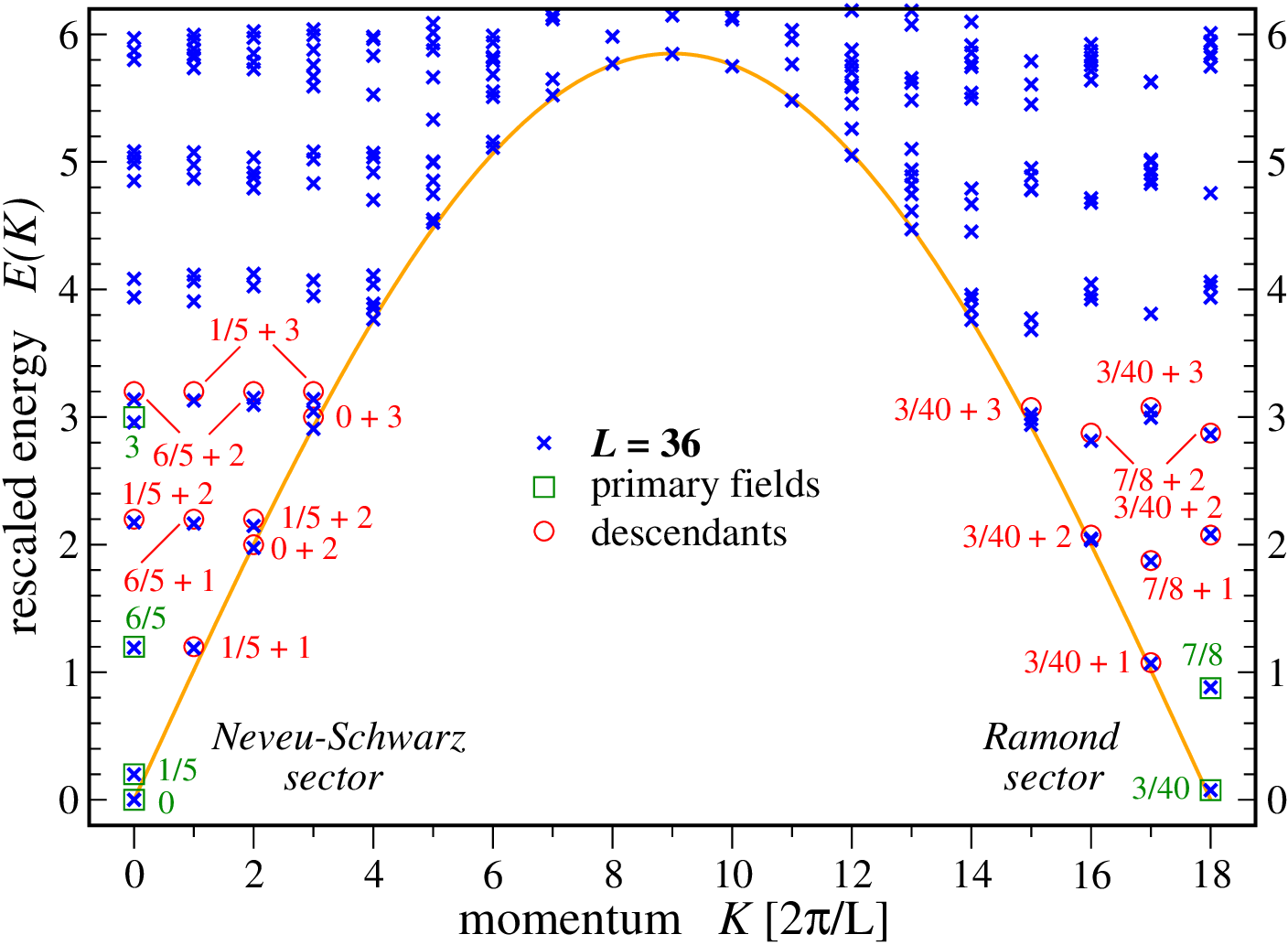}
  \includegraphics[width=86mm]{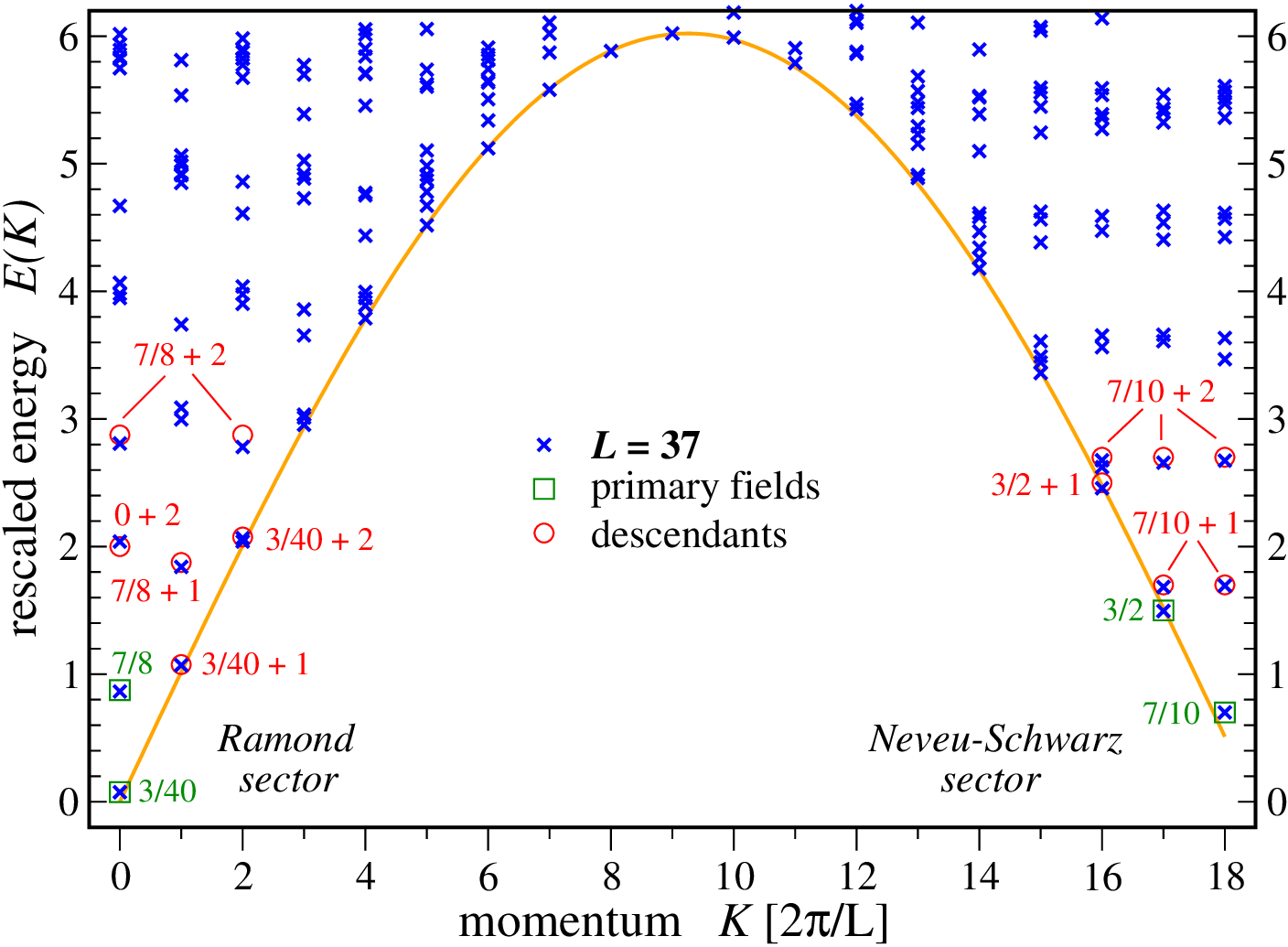}
  \caption{
    Energy spectra for periodic Fibonacci chains of size $L=36$ and $L=37$.
    The spectra have been rescaled and shifted such that the two lowest eigenvalues
    match the conformal field theory assignments.
    The open boxes indicate the positions of the primary fields of the $c=7/10$ conformal
    field theory. The open circles give the positions of multiple descendant fields as indicated. 
    While we find excellent agreement in general, finite-size effects lead to small
    discrepancies for the higher energy states. 
    The solid line is a cosine-fit of the dispersion which serves as a guide to the eye.
  }
  \label{Fig:SpectrumRing}
\end{figure}

\paragraph{Excitation spectra }
We have calculated the excitation spectra of chains up to size $L=37$
with open and periodic boundary conditions using exact diagonalization,
as shown in Fig.~\ref{Fig:SpectrumRing}. 
The numerical results not only confirm the CFT predictions but also
reveal some important details about the correspondence between continuous
fields and microscopic observables. 
In general,  low-energy states on a ring are 
associated with local conformal fields \cite{CardyFSS1984}, 
whose holomorphic and antiholomorphic parts belong to 
representations of the Virasoro algebra,
described by conformal weights $h_L$ and $h_R$.
The  energy levels are given by
\begin{equation}
\label{Eq:SpectrumRing}
E = E_{1}L + \frac{2\pi v}{L} \left(-\frac{c}{12}+h_L+h_R \right) \,,
\end{equation}
corresponding to states with a choice of momenta $K=h_L - h_R$ or $K=h_L - h_R+L/2$ 
in units of $2\pi/L$, where $E_{1}$, $v$ are non-universal constants. 
Here, $h_L= h^{(0)}_L +m_L$ and $h_R= h^{(0)}_R +m_R$, 
where $h^{(0)}_L, h^{(0)}_R$ correspond to weights of `primary' fields
and $m_L$ and $m_R$ are non-negative integers
describing so-called `descendant' fields.
The numerical spectra for even values of $L$ (see the first
plot in Fig.~\ref{Fig:SpectrumRing}) agree with Eq.~(\ref{Eq:SpectrumRing}), 
exhibiting primary fields with $h^{(0)}_L=h^{(0)}_R=0,1/10,3/5,3/2,3/80,7/16$, 
which are conventionally denoted by
$I,\eps,\eps',\eps'',\sigma,\sigma'$, 
respectively\cite{FootnoteOrderOperators}.
The momenta of the last two fields and their descendants 
are near $K=L/2$, as compared to the other four,
indicating that the corresponding microscopic observables have alternating sign 
on the lattice. Such ``staggered'' fields must have nontrivial monodromy with 
respect to a space-time dislocation (i.e., the insertion or removal of a site at some
particular time). 
Such a dislocation is characterized by a chiral $\eps''$ field, say, 
$\eps''_{L}$ \cite{FermionicStressTensor}. 
The $\log$ of the monodromy factor $\exp(2\pi i(h^{\psi\times\eps''}_L-h^{\psi}_L-h^{\eps''}_L))$ 
matches the momenta $K$ in Fig.~\ref{Fig:SpectrumRing}
\cite{FootnoteFusionEpsilon}.
Given this information, we may predict that the states of an odd size ring are
associated with fields of the form $\xi_{L}\eta_{R}$, where
$\eta=\xi\times\eps''$. These include six primary fields, 
$\eps''_{L}$,
$\eps''_{R}$, 
$\eps'_{L} \eps^{\phantom{\prime}}_{R}$,
$\eps^{\phantom{\prime}}_{L}\eps'_{R}$,
$\sigma^{\phantom{\prime}}_{L}\sigma^{\phantom{\prime}}_{R}$,
$\sigma'_{L}\sigma'_{R}$,
 as well as their descendants.
Integrality of the momentum $K$  dictates the choice of $K$ (see below Eq.~(\ref{Eq:SpectrumRing})),
as in Fig.~\ref{Fig:SpectrumRing}.

\begin{table}[t]
  \begin{tabular}{c||c|l||c|l}
  eigenvalue & numerics & $\quad$ CFT & numerics & $\quad$ CFT \\
  & $L=31$ & assignment & $L=32$ & assignment \\ \hline \hline
  0 & 0.10 & $\;\,$ 1/10        & 0      & $\;\,$ 0 \\
  1 & 1.10 & $\;\,$ 1/10 + 1 & 0.60 & $\;\,$ 3/5 \\ \hline
  2 & 1.49 & $\;\,$ 3/2          & 1.60 & $\;\,$ 3/5 + 1\\
  3 & 2.09 & $\;\,$ 1/10 + 2 & 2.02 & $\;\,$ 0 + 2\\
  4 & 2.47 & $\;\,$  3/2 + 1   & 2.58 & $\;\,$ 3/5 + 2 \\
  5 & 3.07 & $\;\,$ 1/10 + 3 & 2.59 & $\;\,$ 3/5 + 2 \\
  6 & 3.11 & $\;\,$ 1/10 + 3 & 3.01 & $\;\,$ 0 + 3\\
  7 & 3.44 & $\;\,$  3/2 + 2   & 3.56 & $\;\,$ 3/5 + 3 \\
  8 & 3.46 &  $\;\,$ 3/2 + 2   & 3.56 & $\;\,$ 3/5 + 3 \\
  \end{tabular}
  \caption{Lowest eigenvalues for open Fibonacci chains of size $L=31$ and $L=32$.
                  The two lowest eigenvalues have been rescaled and shifted such that they 
                  match the conformal field theory assignments.}
  \label{Table:SpectrumChain}
\end{table}

For open boundary conditions the spectra are known to be described by, say,
the holomorphic sector only \cite{Boundary}. To explain the numerical data, we need to
assume that the ends of the chain are charaterized by  a boundary field $\eps$
(or equally well $\eps'$). Thus, for an even number of sites the
spectrum is described by $\eps\times\eps=I+\eps'$ (plus descendants). For an
odd number of sites, this result is to  be modified by fusion with
$\eps''$, yielding $\eps''+\eps$. The numerical low-energy spectra
agree excellently with these predictions as shown in
Table~\ref{Table:SpectrumChain}.


\paragraph{Hidden symmetries }
The critical behavior of our model is not just a peculiarity
of the exact solution but  rather  has topological origin.
In general,  an effective low-energy Lagrangian 
admits  perturbations of the
form $\int \psi(x,\tau)\,dx\,d\tau$, where $\psi$ may be any local field that is
consistent with all applicable symmetries. Such terms are relevant if
$h^{\psi}_L+h^{\psi}_R<2$, in which case they may open a spectral gap or
induce crossover to different critical behavior at large distances.
In the tricritical Ising model, there are four relevant fields: 
$\eps_L\eps_R$, ${\eps'}_L{\eps'}_R$, $\sigma_L\sigma_R$,
${\sigma'}_L{\sigma'}_R$.  Some explanation is in order
as to  why these fields do not appear in the effective Lagrangian
of our model.
The fields $\sigma$ and $\sigma'$ are staggered and thus prohibited by
translational symmetry. Excluding $\eps$ and $\eps'$ requires a
more subtle argument. The Fibonacci ring has a topological symmetry, which
corresponds to adding an extra $\tau$-line parallel to the spine of the fusion
diagram (Fig.~\ref{Fig:FibonacciChain}b) and merging it with the diagram using
the $F$-matrix. We denote this operator by $Y$. 
\begin{equation}
  \langle x_0',\ldots,x_{L-1}'| Y | x_0, \ldots, x_{L-1}\rangle 
  = \prod_{i=0}^{L-1} \left( F^{x'_{i+1}}_{\tau x_i\tau}\right)^{x_i'}_{x_{i+1}} \,,
  \nonumber
\end{equation}
where the identification $L\equiv0$ is used.
We may think of the fusion diagram as a description of a process that generates a 
set of $\tau$-anyons on a circle from the local vacuum. 
Then $Y$ describes another particle moving inside
or outside that circle (or on the circle itself --- before those anyons were
created). The operator $Y$ is sensitive to a possible topological charge
$y=1,\tau$ located at the center of the circle. Thus $Y$ has two eigenvalues,
$S_{y\tau}/S_{y1}=\ph,-\ph^{-1}$. We conjecture that the low-energy states
associated with fields $I,\eps'',\sigma'$ are in the the trivial ($y=1$)
sector, and the fields $\eps',\eps,\sigma$ are in the $y=\tau$ sector. In
fact, the topological fusion algebra (defined by the rule
$\tau\times\tau=1+\tau$) is a quotient of the CFT fusion algebra.

We may imagine that the interaction between the anyons alters the
topological liquid in which the anyons are excitations, producing an annulus of a
different liquid. Some of the local fields correspond to the tunneling of a
$\tau$-anyon between the inner and outer edge of the annulus. Such a
process is actually forbidden as it would change the topological charge
$y$. Thus, only fields in the trivial topological sector  $y=1$ are allowed as
perturbations.  
This excludes
$\psi=\eps_L\eps_R$ and  $\psi={\eps'}_L{\eps'}_R$.


\paragraph{Outlook}
Extensions to chains of anyons in topological liquids
described by  $SU(2)_k$ Chern-Simons theory with $k >3$
corresponding (up to phases) to the non-abelian
statistics of higher members of the Read-Rezayi series
have been mentioned below  Eq.~(\ref{ProjectorsRow}),
but their topological stability is an open issue.
In analogy to quantum spin chains, additional interactions such as dimerization or 
coupling of two Fibonacci  chains in a ladder geometry
should lead to gapped quantum liquids.
Disordered anyonic chains are currently being investigated \cite{Nick}.
For 2D anyonic structures gapless phases  of non-Fermi liquid type
might potentially also emerge.

We thank 
E. Ardonne, 
N. Bonesteel, 
P. Fendley,
C. Nayak, 
G. Refael,
S.~H. Simon, 
and J. Slingerland 
for stimulating discussions.
Some of our numerical simulations were based on the ALPS libraries \cite{ALPS}.



\begin{thebibliography}{99}
\bibitem{FootnoteModularSmatrix}
The matrix $S^{j'}_j $ is the modular S-matrix of $SU(2)_k$ CFT.
\bibitem{uReversedSign}
Here the sign of $u$ in Eq.~(\ref{DefWi})
is reversed~\cite{RSOSmodel}
as compared to the antiferromagnetic case.
\bibitem{FermionicStressTensor} 
$\varepsilon''$ is the superpartner
of the stress tensor in this  $N=1$ superconformal theory \cite{TricriticalIsingModel}.
\bibitem{FootnoteOrderOperators}
The last two fields are  the order operators, i.e. 
those  coupling to applied magnetic fields
in the tricritical Ising model.
\end{thebibliography}
\end{document}